\documentclass[aps,pra,twocolumn,amsmath,nofootinbib]{revtex4-1}
\usepackage{epsfig}
\usepackage[colorlinks,linkcolor=blue,anchorcolor=blue,
            citecolor=blue,urlcolor=blue,
            breaklinks=true]{hyperref}
\usepackage{graphics}
\usepackage{color}
\usepackage{graphicx}
\usepackage{dcolumn}
\usepackage{bm}
\usepackage{cases}
\usepackage{appendix}
\usepackage{ulem}

\newcommand{\tb}{\textbf}

\begin{document}
\author{Yong-Long Wang$^{1,2,3}$}
 \email{wangyonglong@lyu.edu.cn}
\author{Meng-Yun Lai$^{4}$}
\author{Fan Wang$^{4}$}
\author{Hong-Shi Zong$^{4,5,6}$}
\author{Yan-Feng Chen$^{1,3}$}
\email{yfchen@nju.edu.cn}
\address{$^{1}$ National Laboratory of Solid State Microstructures, Department of Materials Science and Engineering, Nanjing University, Nanjing, 210093, China}
\address{$^{2}$ School of Physics and Electronic Engineering, Linyi University, Linyi, 276005, China}
\address{$^{3}$ Collaborative Innovation Center of Advanced Microstructures, Nanjing University, Nanjing 210093, China}
\address{$^{4}$ Department of Physics, Nanjing University, Nanjing 210093, China}
\address{$^{5}$ Joint Center for Particle, Nuclear Physics and Cosmology, Nanjing 210093, China}
\address{$^{6}$ State Key Laboratory of Theoretical Physics, Institute of Theoretical Physics, CAS, Beijing 100190, China}

\title{Geometric effects resulting from square and circular confinements for a particle constrained to a space curve}
\begin{abstract}
Investigating the geometric effects resulting from the detailed behaviors of the confining potential, we consider square and circular confinements to constrain a particle to a space curve. We find a torsion-induced geometric potential and a curvature-induced geometric momentum just in the square case, while a geometric gauge potential solely in the circular case. In the presence of electromagnetic field, a geometrically induced magnetic moment couples with magnetic field as an induced Zeeman coupling only for the circular confinement, also. As spin-orbit interaction is considered, we find some additional terms for the spin-orbit coupling, which are induced not only by torsion, but also curvature. Moreover, in the circular case, the spin also couples with an intrinsic angular momentum, which describes the azimuthal motions mapped on the space curve. As an important conclusion for the thin-layer quantization approach, some substantial geometric effects result from the confinement boundaries. Finally, these results are proved on a helical wire.
\bigskip

\end{abstract}
\maketitle

\section{Introduction}\label{1}
The development of preparing the spatially curved nanowires (for example ~\cite{Deshpande2008, Mourik2012}) has boosted the interest in the quantum physics on a space curve. To study the effective quantum dynamics, a much suitable scheme is the confining potential formalism (CPF) given by Jensen, Koppe, and da Costa~\cite{HJensen1971, Costa1981, Jaffe2003}. In the formalism, a confining potential (CP) is introduced to reduce the dimension. Compensating the reduced motion, a curvature-induced geometric potential (GP) appears in the effective quantum dynamics~\cite{HJensen1971, Costa1981}. The GP can lead to a topological band structure for periodically minimal surfaces~\cite{Aoki2001}, winding-generated bound states for spirally rolled-up nanotubes~\cite{Ortix2010}, reflectionless geometries for bent waveguides~\cite{Campo2014}, the transmission gaps for periodically corrugated thin layers~\cite{Wang2016b} and so on. In addition, a geometric momentum (GM) and a geometric angular momentum~\cite{Wang2017, Liu2011} are defined by curvature. As empirical evidences for the validity of the CPF, the GP~\cite{Szameit2010} has been realized by an optical analog in a topological crystal, and the GM~\cite{Schmidt2015} by the propagation of surface plasmon polaritons on metallic wires.

Further, the CPF has been well discussed in the presence of an external electromagnetic fields~\cite{Ferrari2008, Jensen2009, Ortix2011}, and extended to a particle with spin~\cite{Wang2014}. At the same time, the CPF was used to derive the effective quantum dynamics for spinless particles in a twisted quantum ring~\cite{Taira2010}, a M\"{o}bius ladder~\cite{Sun2009} and a space curve~\cite{Brandt2015}, where the effect of torsion is encoded in an effective magnetic moment, which can cause an effective Zeeman-like coupling~\cite{Sun2009, Brandt2015}, persistent current~\cite{Taira2010} and anomalous phase shift~\cite{Taira2010b}. As spin-orbit coupling (SOC) is considered, the geometric effects become richer~\cite{Ortix2015, Wang2017, Chang2013}, and can generate topological insulating phases~\cite{Ortix2015PRL}. Moreover, pure gauge SOCs appear in the effective quantum dynamics on a ring~\cite{Shikakhwa2017}.

Confining a particle on a space curve embedded in three-dimensional (3D) Euclidean space, Ortix employed double confining potentials~\cite{Ortix2015} to freeze the motions in two reduced dimensions, while Taira and Shima adapted a radial form~\cite{Taira2010} to reduce two dimensions, but just freeze one~\cite{Maraner1995}, the azimuthal motion is preserved. They obtained an additional torsion-induced GP~\cite{Ortix2015} and a geometrically induced gauge potential~\cite{Jaffe2003}, respectively. The torsion-induced GP results from the finite contributions of perturbations, but it does not depend on the specific form of the confining potential and its relative strength~\cite{Ortix2015, Encinosa2005}. The geometric gauge potential is from the nontriviality of a particle constrained in a twisted tube~\cite{Jaffe1999}, and from the geometry of a quantum waveguide~\cite{Schmelcher2014}.

Inspired by those discussions, in this paper we mainly discuss the geometric effects dependent on the detailed behaviors of the CP (its equipotentials around the curve can be squares or circles), for a particle confined to a space curve. We show the torsion-induced GP and the curvature-induced GM just in the square case, and the geometric gauge potential solely in the circular one. With a circular confinement, we obtain an induced Zeeman coupling for the presence of electromagnetic field, and novel terms added to the SOCs by curvature and torsion, and find the remained motions in the effective dynamics as an induced gauge potential and an intrinsic angular momentum. Finally, we consider a concrete example, a helical wire.

\section{Effective dynamics for a particle confined to a space curve}
We will restudy the CPF for a quantum particle confined to a space curve by introducing a CP. The study begins with a space curve $\mathcal{C}$ that is parametrized by $\vec{r}$ depending only on $s$, $s$ is the arc-length. The portion of the immediate neighborhood of $\mathcal{C}$ can be described by
\begin{equation}\label{2-CTube}
\vec{R}(s, q_2, q_3)=\vec{r}+q_2\vec{n}+q_3\vec{b},
\end{equation}
where $\vec{n}$ and $\vec{b}$ are two unit vectors normal and binormal to $\mathcal{C}$, $q_2$ and $q_3$ are the coordinate variables with respect to $\vec{n}$ and $\vec{b}$. The unit vector $\vec{t}$ tangent to $\mathcal{C}$ is defined by $\vec{t}=\partial_s\vec{r}$, where $\partial_s=\frac{\partial}{\partial q_s}$, which is used throughout the paper. With the definition $G_{ij}=\partial_i\vec{R}\cdot\partial_j\vec{R}$ $(i,j=s,2,3)$, the calculation of $G_{ij}$ needs the expression for $\partial_s\vec{t}$, $\partial_s\vec{n}$ and $\partial_s\vec{b}$. The three unit vectors $\vec{t}$, $\vec{n}$ and $\vec{b}$ obey the Frenet-Serret-type equation of motion as they propagate along $s$,
\begin{equation}\label{2-FSE}
\left (
\begin{array}{ccc}
\partial_s\vec{t}\\
\partial_s\vec{n}\\
\partial_s\vec{b}
\end{array}
\right )=
\left (
\begin{array}{ccc}
0 & \kappa & 0 \\
-\kappa & 0 & \tau \\
0 & -\tau & 0
\end{array}
\right )
\left (
\begin{array}{ccc}
\vec{t}\\
\vec{n}\\
\vec{b}
\end{array}
\right ),
\end{equation}
where $\kappa$ and $\tau$ are the curvature and torsion of $\mathcal{C}$, respectively, they may be functions of $s$. Subsequently, $G_{ij}$ can be calculated as follows:
\begin{equation}\label{2-VMetric}
\begin{split}
& G_{11}=(1-\kappa q_2)^2+\tau^2(q_2^2+q_3^2),\\
& G_{12}=G_{21}=-\tau q_3,\quad G_{13}=G_{31}=\tau q_2,\\
& G_{22}=G_{33}=1,\quad G_{23}=G_{32}=0,
\end{split}
\end{equation}
and its determinant is $G=(1-\kappa q_2)^2$. The reduced metric on $\mathcal{C}$ is $g_{ss}=\partial_s\vec{r}\cdot\partial_s\vec{r}=1$ and its determinant is $g=1$. Obviously, $G$ and $g$ satisfy $G=f^2g$ where $f$ is a rescaled factor, $f=1-\kappa q_2$. In terms of Eq.~\eqref{2-VMetric}, the contravariant components of the inverse metric $G^{ij}$ can be derived,
\begin{equation}\label{2-IVMetric}
\begin{split}
& G^{11}=1/f^2,\quad G^{22}=1+\tau^2q_3^2/f^2,\\
& G^{33}=1+\tau^2q_2^2/f^2,\quad G^{12}=G^{21}=\tau q_3/f^2,\\
& G^{13}=G^{31}=-\tau q_2/f^2, G^{23}=G^{32}=-\tau^2q_2q_3/f^2.
\end{split}
\end{equation}

For a particle confined to $\mathcal{C}$, one can employ the CPF to derive the effective Hamiltonian (EH)~\cite{HJensen1971, Costa1981}. In the CPF, the starting point is a quantum dynamics defined in a 3D space, the final aim is an effective quantum dynamics on the one-dimensional (1D) space curve $\mathcal{C}$. Therefore, the motions defined in the plane normal to $\mathcal{C}$ are reduced by introducing a CP. In the spirit of the CPF, for a physical operator $\hat{\rm{F}}$, which does depend purely on derivatives in a 3D Euclidean space, its effective result for a particle confined to $\mathcal{C}$ can be determined by
\begin{equation}\label{2-0-0ER}
\begin{split}
\hat{\rm{F}}_{\rm{eff}}& =\lim_{q_N\to 0}(\hat{\rm{F}} f^{-\frac{1}{2}})-\hat{\rm{F}}_N\\
& =\lim_{q_N\to 0}(1\hat{\rm{F}}f^{-\frac{1}{2}})-\hat{\rm{F}}_N\\
& =\lim_{q_N\to 0}(f^{\frac{1}{2}}\hat{\rm{F}} f^{-\frac{1}{2}})-\hat{\rm{F}}_N,
\end{split}
\end{equation}
where $\hat{\rm{F}}$ is originally defined in an adapted 3D curvilinear coordinate system, $\hat{\rm{F}}_N$ stands for the component operator defined in the normal plane of $\mathcal{C}$, $f^{-\frac{1}{2}}$ is from the normalization condition $\int dsdq_2dq_3\sqrt{g}|f^{\frac{1}{2}}\psi|^2=1$, the third line equality is proved by $\lim_{q_N\to 0}f^{\frac{1}{2}}=1$ and $q_N$ stand for two coordinate variables in the normal plane. With the definition Eq.~\eqref{2-0-0ER}, the curvature-induced GP~\cite{HJensen1971, Costa1981} and GM~\cite{Liu2011, Wang2017} can be well obtained. As da Costa said, the GP is independent of the detailed behavior of the CP (its equipotentials around $\mathcal{C}$ can be circles, squares, elipses, etc.)~\cite{Costa1981}. The independence is confirmed by limiting $q_N\to 0$ in Eq.~\eqref{2-0-0ER}. It is essentially important in separating the tangent component from the normal ones analytically that the limitation accomplishes the disappearance of the mixed terms of tangent dimension and normal ones.

We notice that the torsion-induced GP given by Ortix~\cite{Ortix2015} and the geometrically induced magnetic moment by Brandt and S\'{a}nchez-Monroy~\cite{Brandt2015} are both from the first-order perturbations. Those results can not be obtained through Eq.~\eqref{2-0-0ER}. Inspired by the extents of perturbation theories carried out in Ref.~\cite{Ortix2015, Brandt2015}, we investigate the geometric effects resulting from the confinement boundaries of the GP. In the case, the calculation procedure of CPF has to be reconsidered. The particle is initially proposed to occur in a quantum state that is $|\chi_s\rangle\otimes|\chi_N\rangle$ ($|\chi_s\rangle$ denotes an eigenstate in the $s$ dimension and $|\chi_N\rangle$ stands for that in the normal plane of $\mathcal{C}$). Because the particle is initially described by a 3D quantum equation, and the ultimate aim of CPF is the analytical separation of the $s$-dimensional quantum dynamics and the normal component. The CP increases the excited energies of confined degree of freedom far beyond those of unconfined one. As a result, the motion in confined dimension is frozen so that the particle solely exists in its ground state. Therefore, the formula Eq.~\eqref{2-0-0ER} should be reexpressed as
\begin{equation}\label{2-0-1ER}
\begin{split}
\hat{\rm{F}}_{\rm{eff}}&=\lim_{\varepsilon_c\to 0}\langle\chi_{0_c}|f^{\frac{1}{2}}\hat{\rm{F}}f^{-\frac{1}{2}} -\hat{\rm{F}}_N|\chi_{0_c}\rangle\\
&=\langle\chi_{0_c}|f^{\frac{1}{2}}\hat{\rm{F}}f^{-\frac{1}{2}} -\hat{\rm{F}}_N|\chi_{0_c}\rangle_0,
\end{split}
\end{equation}
where $\varepsilon_{c}$ describes the scale size of the confined degree of freedom, $|\chi_{0_c}\rangle$ denotes the confined ground state and the subscript $0$ in $\langle\cdots\rangle_0$ stands for the limit $\varepsilon_c\to 0$. The validity of the formula~\eqref{2-0-1ER} is protected by the two aims of the CPF~\cite{Wang2016}: to separate the quantum equation tangent to $\mathcal{C}$ from the normal components analytically, to preserve the normal motions in the EH as much as possible. It is worthwhile to notice that $|\chi_{0_c}\rangle$ is the ground state of confined dimension, not the reduced one. Compared to Eq.~\eqref{2-0-0ER}, Eq.~\eqref{2-0-1ER} adds one step to bring the finite contributions of the original operator defined in the ground state $|\chi_{0_c}\rangle$ back into the effective result. The specific form of $|\chi_{0_c}\rangle$ is eventually determined by the confinement boundaries, and the added contributions result from them naturally.

\subsection{Spinless particle}
A free spinless particle can be described by the Hamiltonian $\hat{\rm{H}}_0$,
\begin{equation}\label{2-FHamilton}
\hat{\rm{H}}_0=-\frac{\hbar^2}{2m}\nabla^2,
\end{equation}
where $\hbar$ is the Plank constant divided by $2\pi$ and $m$ is the effective mass of the particle. Once a CP denoted as $V_c$ is introduced to confine the particle to $\mathcal{C}$, $\hat{\rm{H}}_0$ would be replaced by
\begin{equation}\label{2-VHamilton}
\hat{\rm{H}}_V=\hat{\rm{H}}_0+V_c.
\end{equation}

Originally, da Costa gave $V_c$~\cite{Costa1981} in the following form
\begin{equation}\label{2-ConfPot}
V_c(q_{2,3})=
\begin{cases}
0, \quad q_2=0,q_3=0, \\
\infty, \quad q_2\neq 0, q_3\neq 0.
\end{cases}
\end{equation}
The considered realization, $V_c$ should be reexpressed as
\begin{equation}\label{2-InfWell}
V_c(q_{2,3})=\lim_{\varepsilon_{2,3}\to 0}
\begin{cases}
0, \quad |q_2|\leq\varepsilon_2,\quad|q_3|\leq\varepsilon_3, \\
\infty, \quad |q_2|>\varepsilon_2, \quad |q_3|>\varepsilon_3,
\end{cases}
\end{equation}
where $\varepsilon_2$ and $\varepsilon_3$ describe the scale sizes of the plane normal to $\mathcal{C}$. More specifically, double CPs can be chosen as
\begin{equation}\label{2-HarmOsc}
\begin{split}
&V_c(q_2)=\lim_{w\to\infty}\frac{1}{2}m w^2q_2^2,\\
&V_c(q_3)=\lim_{w\to\infty}\frac{1}{2}m w^2q_3^2,
\end{split}
\end{equation}
where $w$ denotes a harmonic frequency, where the independence of Eq.~\eqref{2-0-1ER} on the specific form of the CP and its relative strength is considered~\cite{Ortix2015, Encinosa2005}.
\begin{figure}[htbp]
\centering
\includegraphics[width=0.44\textwidth]{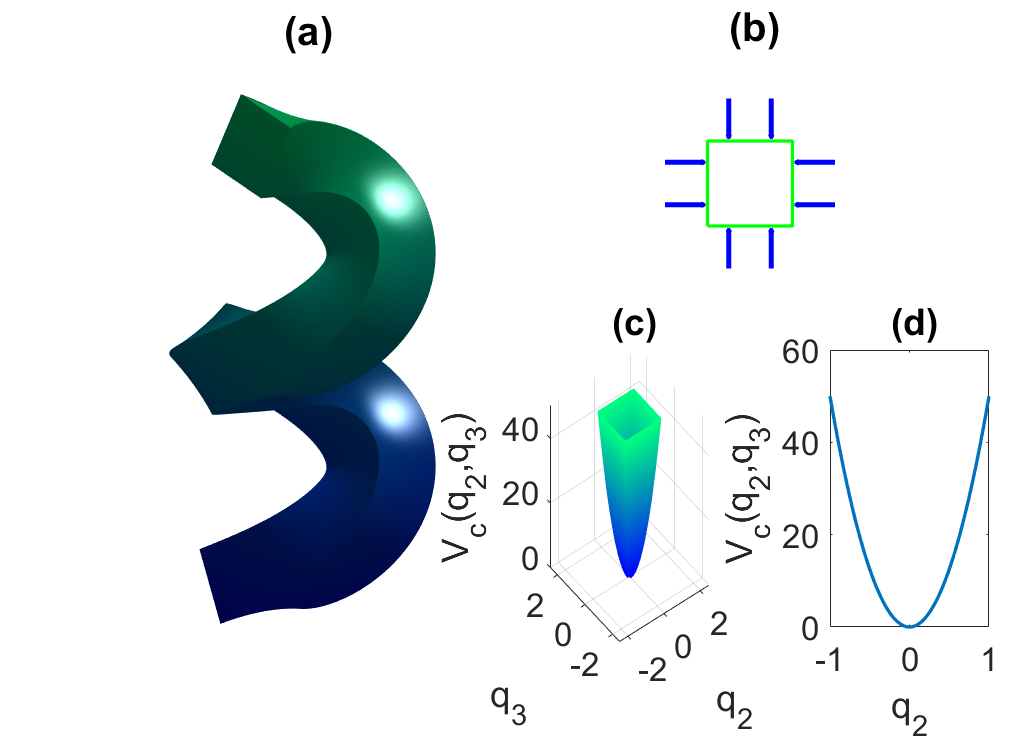}
\caption{\footnotesize Sketches of the square case. (a) A twisted tube with square cross section. (b) A cross section with squeezing forces sketched. (c) The double confining potential vs $(q_2, q_3)$. (d) The confining potential vs $q_2$. The confining potential is regarded as a dimensionless quantity, because its unit is being taken as the excited energy in $s$ dimension.}\label{Square}
\end{figure}

By virtue of the analytical separation of the tangent motion and that in the normal plane, and the CP Eq.~\eqref{2-HarmOsc}, one can directly give the Hamiltonian in the plane normal to $\mathcal{C}$ as $\hat{\rm{H}}_N=\hat{\rm{H}}_n+\hat{\rm{H}}_b$. Introduced as a square confinement as sketched in Fig.~\ref{Square}, the normal component $\hat{\rm{H}}_n$ and binormal one $\hat{\rm{H}}_b$ can be written as
\begin{equation}\label{2-1-nH}
\hat{\rm{H}}_n=-\frac{\hbar^2}{2m}\partial_2^2+\frac{1}{2}mw^2q_2^2,
\end{equation}
and
\begin{equation}\label{2-1-bH}
\hat{\rm{H}}_b=-\frac{\hbar^2}{2m}\partial_3^2+\frac{1}{2}mw^2q_3^2,
\end{equation}
respectively. The simplicity of Eqs.~\eqref{2-1-nH} and ~\eqref{2-1-bH} is eventually determined by the extreme strength of the CP, $w\to\infty$. All the other terms depending on $q_{2,3}$ or $\partial_{2,3}$ can be taken as perturbations. The quantum mechanical problem, 1D harmonic oscillator, is exactly solvable and does not involve complicated calculation. The normal and binormal ground states are
\begin{equation}\label{2-NWF}
|\chi_{0_n}\rangle=\alpha^{1/2}\pi^{-1/4}e^{-(\alpha q_2)^2/2},
\end{equation}
and
\begin{equation}\label{2-BWF}
|\chi_{0_b}\rangle=\alpha^{1/2}\pi^{-1/4}e^{-(\alpha q_3)^2/2},
\end{equation}
respectively, where $\alpha=\sqrt{mw/\hbar}$. The corresponding zero-point energies are $E_{0_n}=E_{0_b}=\hbar w/2$.

In terms of Eqs.~\eqref{2-0-1ER}, ~\eqref{2-VHamilton}, ~\eqref{2-NWF} and ~\eqref{2-BWF}, the EH can be calculated,
\begin{equation}\label{2-1SEH}
\begin{split}
\hat{\rm{H}}_{\rm{eff}}&=\langle\chi_{0_n, 0_b}|f^{\frac{1}{2}}\hat{\rm{H}}_Vf^{-\frac{1}{2}}-\hat{\rm{H}}_N|\chi_{0_n, 0_b}\rangle_0\\
&=\lim_{w\to\infty}\langle\chi_{0_n, 0_b}|f^{\frac{1}{2}}\hat{\rm{H}}_Vf^{-\frac{1}{2}}-\hat{\rm{H}}_N|\chi_{0_n, 0_b}\rangle\\
&=\frac{1}{2m}\hat{p}_s^2-\frac{\hbar^2}{8m}\kappa^2 -\frac{\hbar^2}{4m}\tau^2,
\end{split}
\end{equation}
where $|\chi_{0_n, 0_b}\rangle=|\chi_{0_n}\rangle\otimes|\chi_{0_b}\rangle$, $\hat{p}_s=-i\hbar\partial_s$ is a kinematic momentum operator, $-\frac{\hbar^2}{8m}\kappa^2$ is the well-known GP induced by curvature and $-\frac{\hbar^2}{4m}\tau^2$ is an additional GP induced by torsion. This result is in full agreement with that in Ref.~\cite{Ortix2015} without spin. It is easy to prove that the curvature-induced GP can be also obtained by Eq.~\eqref{2-0-0ER}, but the torsion-induced GP can not, which results from the finite expectations of the expression, that is
\begin{equation}\nonumber
\frac{\hbar^2}{2m}\tau^2\langle\chi_{0_n,0_b}|(q_2\partial_2+ q_3\partial_3+2q_2 q_3\partial_2\partial_3)|\chi_{0_n,0_b}\rangle_0.
\end{equation}
In the previous calculation process, the limit $\varepsilon\to 0$ is fully equivalent to $w\to\infty$. The reason is the ground state width being $\varepsilon=\sqrt{\hbar/(mw)}$.

\begin{figure}[htbp]
\centering
\includegraphics[width=0.44\textwidth]{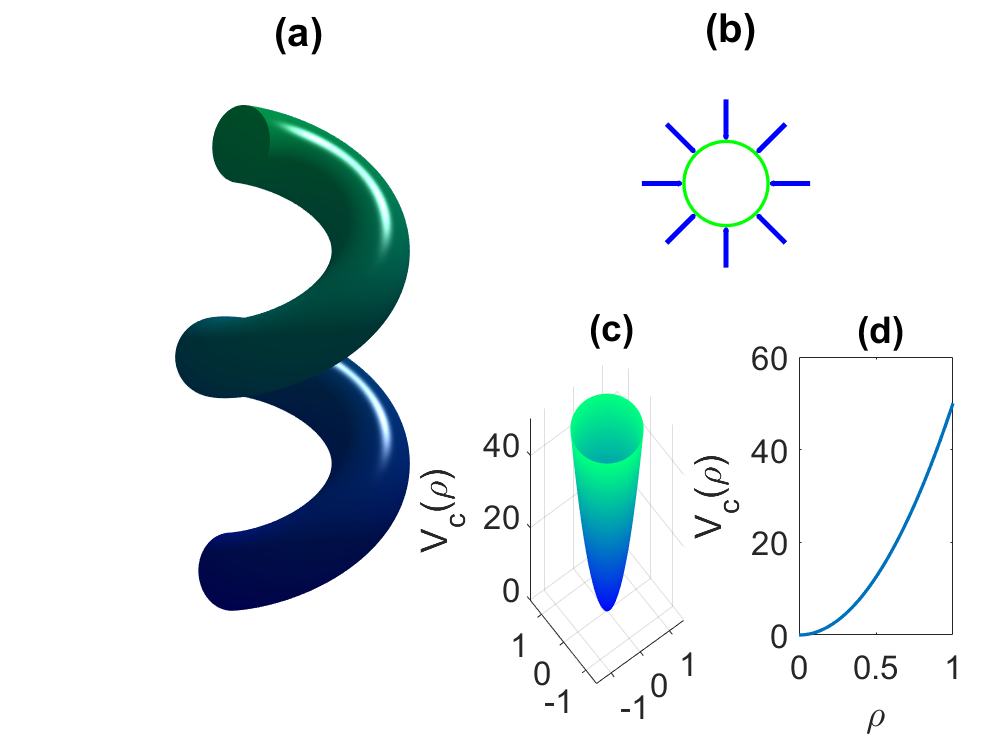}
\caption{\footnotesize Sketches of the circular case. (a) A twisted tube. (b) A cross section with squeezing forces sketched. (c) The radial confining potential vs the variables $(\rho, \theta)$ sketched. (d) The radial confining potential vs $\rho$ sketched. The confining potential is regarded as a dimensionless quantity, because its unit is being taken as the excited energy in $s$ dimension.}\label{Circular}
\end{figure}
In the circular case sketched in Fig.~\ref{Circular}, the CP ~\eqref{2-HarmOsc} can be rewritten as
\begin{equation}\label{2-RHarmOsc}
V_c(\rho)=\lim_{w\to\infty}\frac{1}{2}mw^2\rho^2,
\end{equation}
where $w$ is a harmonic frequency. It is a striking difference that the square confinement freezes two degrees of freedom, but only one in the circular case, in which the azimuthal motion is remained. It is obvious that $V_c(\rho)$ possesses a SO(2) symmetry. The angular momentum $\hat{\rm{L}}_s$, therefore, commutes with the Hamiltonian $\hat{\rm{H}}_N$. $\hat{\rm{L}}_s$ can be expressed by
\begin{equation}\label{2-AngularMomentum}
\hat{\rm{L}}_s=i\hbar(q_3\partial_2-q_2\partial_3)=-i\hbar\partial_{\theta},
\end{equation}
where $\partial_2=\partial/\partial q_2$, $\partial_3=\partial/\partial q_3$ and $\theta$ denotes the azimuthal variable in the normal plane of $\mathcal{C}$. In a local polar coordinate system, $\hat{\rm{H}}_N$ reads
\begin{equation}\label{2-RHamilton0}
\hat{\rm{H}}_N=-\frac{\hbar^2}{2m}\frac{1}{\rho}\partial_{\rho}\rho\partial_{\rho} +\frac{1}{2m\rho^2}\hat{\rm{L}}_s^2+\frac{1}{2}mw^2\rho^2.
\end{equation}
Since $[\hat{\rm{H}}_N,\hat{\rm{L}}_s]=0$, there are a set of states which are common eigenstates of $\hat{\rm{H}}_N$ and $\hat{\rm{L}}_s$. Their eigen-equations are
\begin{equation}\label{2-CSE0}
\hat{\rm{H}}_N|\chi_{n,l}\rangle =E_{n,l}|\chi_{n,l}\rangle,
\end{equation}
and
\begin{equation}\label{2-AngMom0}
\hat{\rm{L}}_s|\chi_{n,l}\rangle =l\hbar|\chi_{n,l}\rangle,
\end{equation}
respectively, where $E_{n,l}$ is the eigenvalue of $\hat{\rm{H}}_N$, $l\hbar$ is the eigenvalue of $\hat{\rm{L}}_s$ and $|\chi_{n,l}\rangle$ denotes a common eigenstate, wherein $n$ is a radial quantum number and $l$ is an azimuthal quantum number. The CP~\eqref{2-RHarmOsc} freezes the radial degree of freedom so that $n$ has a sole value $n=0$. $l$ without difference still has $0,\pm1,\pm2, ...$. Fortunately, the ground state $|\chi_{0,l}\rangle$ can be solved in the exact form
\begin{equation}\label{2-CWF}
|\chi_{0,l}\rangle=A e^{il\theta}(\beta\rho)^{|l|} e^{-\beta^2\rho^2/2},
\end{equation}
where $A$ is a normalized constant with $A=\sqrt{\frac{2^{|l|+1}\beta} {\sqrt{\pi}(2|l|-1)!!}} $ and $\beta=\sqrt{\frac{mw}{\hbar}}$, wherein "$!!$" denotes double factorial. This result may open an access to discuss electrons with an intrinsic orbital angular momentum in a 1D system~\cite{Stephen2017, Iwo2017}.

According to Eqs.~\eqref{2-0-1ER}, \eqref{2-VHamilton}, \eqref{2-CSE0} and \eqref{2-CWF}, the EH can be deduced,
\begin{equation}\label{2-1CEH}
\begin{split}
\hat{\rm{H}}_{\rm{eff}}& =\langle\chi_{0,l}| f^{\frac{1}{2}}\hat{\rm{H}}_Vf^{-\frac{1}{2}} -\hat{\rm{H}}_N |\chi_{0,l}\rangle_0\\
&=\frac{1}{2m}(\hat{p}_s-l\hbar\tau)^2-\frac{\hbar^2}{8m}\kappa^2.
\end{split}
\end{equation}
In the calculation process, the equivalence between $\varepsilon\to 0$ and $w\to\infty$ is also considered. Compared to Eq.~\eqref{2-1SEH}, we find a striking distinction that an additional term $-l\hbar\tau$ appears in Eq.~\eqref{2-1CEH}, while the torsion-induced GP vanishes. Further, the difference results from the confinement boundaries. The square confinement breaks the SO(2) symmetry that leads to the presence of the torsion-induced GP. The circle CP, however, holds the SO(2) symmetry to provide an additional term $-l\hbar\tau$ to the $s$ momentum. As an important conclusion, the geometric effects can be induced by the geometry of curve, and can be contributed by the confinement boundaries.

To learn the gauge structure of the EH~\eqref{2-1CEH}, we reconsider the appearance $-l\hbar \tau$. Here $\tau$ is the torsion of $\mathcal{C}$ and it can be given by the normal fundamental form $A_s^{23}=-A_s^{32}=\tau$, which is defined by $A_s^{ab}=\vec{n}_a\cdot\partial_s\vec{n}_b$ $(a,b=2,3)$. And $l\hbar$ can be given by
\begin{equation}
l\hbar=\langle\chi_{0,l}| \hat{\rm{L}}_s|\chi_{0,l}\rangle_0
=\langle\chi_{0,l}|\hat{\rm{L}}_{23}|\chi_{0,l}\rangle_0,
\end{equation}
where $\hat{\rm{L}}_{23}=i\hbar(q_3\partial_2-q_2\partial_3)$.

Under a point-dependent rotation $\mathcal{R}$~\cite{Jaffe2003}, $\vec{n}_2$ and $\vec{n}_3$ transform as
\begin{equation}\label{2-GTrans1}
\left (
\begin{array}{c}
\vec{n}_2^{\prime}\\
\vec{n}_3^{\prime}
\end{array}
\right )=\mathcal{R}\left(
\begin{array}{c}
\vec{n}_2\\
\vec{n}_3
\end{array}
\right )
\end{equation}
where $\mathcal{R}=e^{i\theta_{ab}\hat{\rm{L}}_{ab}}$, $A^{ab}_s$ transforms as an SO(2) gauge connection
\begin{equation}\label{2-GTrans2}
A_s^{ab}\rightarrow \mathcal{R}^{ac}A_s^{cd}\mathcal{R}^{db} +\mathcal{R}^{ac}\partial_s\mathcal{R}^{cb},
\end{equation}
and then $|\chi_s\rangle$ and $A_s$ transform as
\begin{equation}\label{2-GTrans3}
\begin{split}
& |\chi_s\rangle\rightarrow \mathcal{V}|\chi_s\rangle,\\
& A_s\rightarrow \mathcal{V}A_s\mathcal{V}^T+\mathcal{V}\partial_s\mathcal{V}^T,
\end{split}
\end{equation}
where $A_s=A_s^{ab}l_{ab}$, $\mathcal{V}=\langle\mathcal{R} \rangle_0=e^{i\theta_{ab}l_{ab}}$ with $l_{ab}=\langle\hat{\rm{L}}_{ab}\rangle_0$ and the $s$-independence of $|\chi_{0,l}\rangle$ is considered. It is apparent that $A_s$ is only present when $|\chi_{0, l}\rangle$ is the nontrivial representation of SO(2) with $l_{ab}\neq 0$. For the CP with abelian $\rm{SO}(2)\cong \rm{U}(1)$ invariance the U(1)-induced gauge potential $A_s$ cannot be transformed away~\cite{Jaffe2003}. Therefore it is physical and measurable, not a pure mathematical connection~\cite{Chen2008, Wang2015}. In other words, the geometrical torsion provides a platform to measure the induced gauge potential.

Due to $A_s$ being induced by the torsion $\tau$, it can be named as the geometric gauge potential~\cite{Jaffe2003}, and denoted by $A_g$, $A_g=l\hbar \tau$. Thus the EH~\eqref{2-1CEH} can be rewritten as
\begin{equation}\label{2-1SCEH}
\hat{\rm{H}}_{\rm{eff}}=\frac{1}{2m}(\hat{p}_s-A_g)^2-\frac{\hbar^2}{8m}\kappa^2.
\end{equation}
The appearance of $A_g$ preserves the U(1) gauge invariance of the EH. This result would provide a way to generate an artificial gauge field by designing the geometry of nanodevice.  It is worthwhile to notice that the gauge potential is global for the torsion $\tau$ being a constant, local for a function of $s$.

A momentum operator $\vec{\rm{p}}$ in the $(\vec{t}, \vec{n}, \vec{b})$ coordinate system is
\begin{equation}\label{2-1-SM}
\vec{\rm{p}}=-i\hbar(\vec{t}\frac{1}{\sqrt{G_{11}}}\partial_s +\vec{n}\partial_2 +\vec{b}\partial_3),
\end{equation}
where $G_{22}=G_{33}=1$ is considered. Using Eq.~\eqref{2-0-1ER}, the effective momentum for the square confinement can be deduced,
\begin{equation}\label{2-1-SEM}
\begin{split}
\vec{\rm{p}}_{\rm{eff}}&=\langle\chi_{0_n,0_b}|f^{\frac{1}{2}}\vec{\rm{p}}f^{-\frac{1}{2}} -\vec{\rm{p}}_N|\chi_{0_n,0_b}\rangle_0\\
&=\vec{\rm{p}}_s+\vec{\rm{p}}_g=-i\hbar(\vec{t}\partial_s-\frac{1}{2}\kappa \vec{n}),
\end{split}
\end{equation}
where $\vec{\rm{p}}_s$ is the $s$ momentum operator vector, $\vec{\rm{p}}_s=-i\hbar\vec{t}\partial_s$, and $\vec{\rm{p}}_g$ is a term induced by curvature that is named a GM, $\vec{\rm{p}}_g=i\hbar\frac{\kappa}{2}\vec{n}$.

In the circular case, the momentum operator can be written,
\begin{equation}\label{2-1-CM}
\vec{\rm{p}}=-i\hbar(\vec{t}\frac{1}{\sqrt{G_{11}}}\partial_s +\vec{e}_{\theta}\frac{1}{\rho}\partial_{\theta} +\vec{e}_{\rho}\partial_{\rho}).
\end{equation}
where $G_{11}=(1-\kappa \rho\cos\theta)^2+\tau^2\rho^2$, $\rho=\sqrt{q_2^2+q_3^2}$ and $\theta=\arctan(q_3/q_2)$. With the circular confinement, we obtain a trivial effective momentum $\vec{\rm{p}}_{\rm{eff}}$, that is
\begin{equation}\label{2-1-CEM}
\vec{\rm{p}}_{eff}=\langle\chi_{0,l}|f^{\frac{1}{2}}\vec{\rm{p}}f^{-\frac{1}{2}} -\vec{\rm{p}}_N|\chi_{0,l}\rangle_0
=\vec{\rm{p}}_s,
\end{equation}
where $f=1-\kappa\rho\cos\theta$. Here the GM is not displayed.

\subsection{Charged particle without spin}
A charged particle in electromagnetic field can be described by
\begin{equation}\label{2-2H}
\hat{\rm{H}}=-\frac{\hbar^2}{2m}\frac{1}{\sqrt{G}}\mathcal{D}_{\mu}\sqrt{G} G^{\mu\nu}\mathcal{D}_{\nu}-eA_0,
\end{equation}
where $A_0=-\phi$, $\phi$ is the scalar potential, $e$ is unit charge, $\mathcal{D}_{\mu}=\partial_{\mu}-\frac{ie}{\hbar}A_{\mu}$ are the gauge covariant spatial derivatives, wherein $A_{\mu}$ are the components of the vector potential $\vec{A}$, $\mu,\nu=1,2,3$ denote three curvilinear coordinates, and $G$ and $G^{\mu\nu}$ are the determinant and inverse of the metric $G_{\mu\nu}$, respectively.

Repeating the previous calculations, with a square CP we can obtain the EH for a charged particle constrained to $\mathcal{C}$ in electromagnetic field, that is
\begin{equation}\label{2-2SEH}
\hat{\rm{H}}_{\rm{eff}}=\frac{1}{2m}(\hat{p}_s-e\bar{A}_s)^2
-\frac{\hbar^2}{8m}\kappa^2 -\frac{\hbar^2}{4m}\tau^2-eA_0,
\end{equation}
where $\bar{A}_s$ is the $s$ component of the electromagnetic potential evaluated at $q_{2,3}=0$, $\bar{A}_s=A_s(s, q_{2,3})|_{q_{2,3}=0}$~\cite{Brandt2015}. The presence of electromagnetic field can not lead any new geometric effects except the GPs induced by curvature and torsion.

For the circular CP, the EH is also obtained,
\begin{equation}\label{2-2CEH}
\hat{\rm{H}}_{\rm{eff}}=\frac{1}{2m}(\hat{p}_s-A_g -e\bar{A}_s)^2-\frac{\hbar^2}{8m}\kappa^2+B_s\mu_g-eA_0,
\end{equation}
where $A_g=l\hbar\tau$ is the geometric gauge potential, $\bar{A}_s=A_s(s,0)$, $B_s$ is the magnetic field tangent to $\mathcal{C}$ defined by $B_s=\bar{F}_{23}=(\partial_2A_3-\partial_3A_2)|_{q_{2,3}\to 0}$ and $\mu_g=l\mu$ can be taken as a geometrically induced magnetic moment, wherein $\mu=-\frac{\hbar e}{2m}$ and $l$ is an quantum number of angular momentum. For a particle confined to a space curve, the radial confinement leads to the azimuthal motions mapped on the 1D space curve. We notice that $\mu_g$ does not depend on the curvature and torsion, and then does not depend on the position of particle on $\mathcal{C}$. As an intrinsic angular momentum that is an angular momentum independent of the position~\cite{Neil2002, Zhang2005}, here $\mu_g$ can be taken as an intrinsic magnetic moment.

\subsection{Particle with SOC}
As a particle is considered SOC, its Hamiltonian can be described by
\begin{equation}\label{2-3H}
\hat{\rm{H}}=-\frac{\hbar^2}{2m}\frac{1}{\sqrt{G}} \partial_{\mu}\sqrt{G}G^{\mu\nu}\partial_{\nu}-i\hbar\frac{1}{\sqrt{G}}\varepsilon^{\mu\nu\lambda} \alpha_{\mu}\Sigma_{\nu}\partial_{\lambda},
\end{equation}
where $\mu,\nu,\lambda=s,2,3$, $\varepsilon^{\mu\nu\lambda}$ is the Levi-Civita symbol, $\alpha_{\mu}$ are the coefficient constants of the SOC, $\Sigma_{\nu}$ denote three induced Pauli matrices, $\Sigma_{\nu}=\vec{\tb{e}}^{i}_{\nu}\sigma_{i}$~\cite{Dick2006, Wang2014}, wherein $\sigma_{i}$ $(i=x,y,z)$ are the usual Pauli matrices.

Following the above calculation procedure, the EH for the square CP is obtained,
\begin{equation}\label{2-3SEH}
\begin{split}
\hat{\rm{H}}_{\rm{eff}}=&\frac{\hat{p}_s^2}{2m}-\frac{\hbar^2}{8m}\kappa^2 -\frac{\hbar^2}{4m}\tau^2-i\hbar(\alpha_s\sigma_b-\alpha_b\sigma_s) \frac{\kappa}{2} \\
& +\alpha_n(\sigma_b\hat{p}_s+i\hbar\sigma_n\frac{\tau}{2}) -\alpha_b (\sigma_n\hat{p}_s-i\hbar\sigma_b\frac{\tau}{2}).
\end{split}
\end{equation}
This result shows that the additional terms to the SOC are induced not only by $\tau$~\cite{Ortix2015}, but also $\kappa$.

In the circular case, $\Sigma_s=\sigma_s(1-\kappa\rho\cos\theta) +\sigma_b\tau\rho\cos\theta -\sigma_n\tau\rho\sin\theta$, $\Sigma_{\rho}=\sigma_n\cos\theta+\sigma_b\sin\theta$, and $\Sigma_{\theta}=-\sigma_n\sin\theta+\sigma_b\cos\theta$ in the local polar coordinates system, respectively, the EH is reobtained,
\begin{equation}\label{2-ESPESO2}
\begin{split}
\hat{\rm{H}}_{\rm{eff}}&= \frac{1}{2m}(\hat{p}_s-A_g)^2-\frac{\hbar^2}{8m} \kappa^2 \\
& +i\hbar(\alpha_b\sigma_s-3\alpha_s\sigma_b)\frac{\kappa}{8} -l\hbar(\alpha_s\sigma_n-3\alpha_n\sigma_s)\frac{\kappa}{4}\\ &+\alpha_n[\sigma_b(\hat{p}_s -\frac{A_g}{2}) +i\hbar\sigma_n\frac{\tau}{4}]\\
& -\alpha_b [\sigma_n(\hat{p}_s-\frac{A_g}{2})-i\hbar\sigma_b\frac{\tau}{4}].
\end{split}
\end{equation}
Obviously, there appear new terms induced by $\kappa$ and $\tau$ added to SOCs. When $\tau=0$, the intrinsic angular momentum $l\hbar$ cannot be completely eliminated from Eq.~\eqref{2-ESPESO2}. The reason is that the azimuthal motion also couples with spin through $\kappa$. As a consequence, the circular confinement provides richer geometric effects for a particle constrained to a space curve.

\section{Helical Wire}
As an example, we consider a right-handed helical wire~\cite{Ortix2015, Qi2009} described by $\vec{r}=(r\cos\theta, r\sin\theta, c\theta)$,
where $r$ and $2\pi c$ are the radius and pitch of the helix, and $\theta$ denotes the azimuthal angle that is defined by $\theta=s/L$, wherein $L=\sqrt{r^2+c^2}$ and $s$ is the arc-length. And then we can obtain the curvature $\kappa=r/L^2$, the torsion $\tau=c/L^2$ and the tripod of orthonormal vectors
$\vec{t}=(-\kappa\sin\theta,\kappa\cos\theta,\tau)$,
$\vec{n}=(-\cos\theta,-\sin\theta,0)$,
$\vec{b}=(\tau\sin\theta,-\tau\cos\theta,\kappa)$.

In the circular case, the nontrivial ground states are all doubly degenerate except the trivial one, $l=0$. With respect to the ground states, the eigenvalues of energy in the plane normal to the helical wire can be described by
\begin{equation}
E_{0,l}=2(l-\frac{1}{2})(l+\frac{1}{2})\hbar w, \quad l=0,\pm 1, \pm 2,\cdots.
\end{equation}
And thus the effective energy spectrum can be derived as
\begin{equation}
E_{\rm{eff}\pm}=\frac{1}{2m}(p_s\mp l\hbar\tau)^2-\frac{\hbar^2\kappa^2}{8m}.
\end{equation}

In the presence of an externally applied electromagnetic field, the eigenenergy in the normal plane should be reexpressed by
\begin{equation}
E_{0,l}=2(l-\frac{1}{2})(l+\frac{1}{2})\hbar w+B_s\mu_g.
\end{equation}
With the limit $w\to\infty$, $B_s\mu_g$ is a perturbation, which could be preserved in the effective dynamics. For the double degenerate ground state with $l=\pm 1$, the effective energy spectrum can be obtained,
\begin{equation}
E_{\rm{eff}\pm}=\frac{1}{2m}(p_s\mp\hbar\frac{c}{L^2}-e\bar{A}_s)^2 -\frac{\hbar^2r^2}{8mL^4}\pm B_s\mu_g.
\end{equation}
Notice that the azimuthal motion appears in the normal component, and also induces a gauge potential and a magnetic moment in the effective energy.

As the SOC is considered, using a trial spinorial wavefunction
$|\chi_s\rangle\otimes|s\rangle=\chi(s)[e^{-\frac{1}{2}i\theta},
e^{\frac{1}{2}i\theta}]^T$,
the effective energy spectrum with $l=\pm1$ can be obtained,
\begin{equation}
\begin{split}
E_{\rm{eff}\pm}=& \frac{1}{2m}(p_s\mp \hbar\frac{c}{L^2})^2-\frac{\hbar^2}{8m}\frac{r^2}{L^4}\pm \hbar\alpha_s\frac{r}{2L^2}\\
& -i\hbar\alpha_n\frac{c}{2L^2}+2\alpha_b(p_s\mp \hbar\frac{c}{2L^2}).
\end{split}
\end{equation}

These results mostly do not appear in the square case, which clearly illustrate that the preserved motion of reduced dimension is as important as the geometry of the helical wire in the geometric effects. The preservation is determined by the circular confinement. In other words, the detailed behaviors of the CP play an important role in the geometric effects.

\section{Conclusions and discussions}
In the spirit of the CPF, we first reconsider the fundamental calculation expression, the effective dynamics should be defined by the ground state of the confined dimension. It is worthwhile to notice that the number of the confined dimensions may be not identical to that of the reduced ones. To distinguish them, we deduce the effective dynamics for a particle confined to a space curved embedded in 3D space by using square and circular CPs. In the square case, the confined dimensions are identical to the reduced ones. However, one dimension is confined, but two is reduced for the circular confinement.

We demonstrate that the curvature-induced GP does not depend on the detailed behaviors of the CP~\cite{Costa1981}, which as predicted identically appears in the two cases. However, we find that the torsion-induced GP, the geometric momentum and the geometric gauge potential do depend on the confinement boundaries. The torsion-induced PG and the geometric momentum appear only in the square case, the geometric gauge potential is merely displayed in the circular case. In the presence of electromagnetic field, we show the induced magnetic moment solely for the circularly squeezing. As SOC is considered, the square confinement provides the curvature and torsion induced SOCs, the circular CP provides the action of azimuthal motion on SOC. Distinctly, the torsion induced SOCs in the circular case are exactly half of those in the square case.

On the basis of the discussions in the present paper, the fundamental framework of the CPF generally consists of three steps: (1) solve the ground state of the confined degree of freedom, (2) calculate the rescaled operator by averaging over the ground state, and (3) limit the scale size of confined dimension. As evidence, the effective dynamics can be in general defined in terms of the ground state of the confined dimension. This definition provides us a powerful approach to discuss the generation of an artificial gauge field and the manipulation of spin-transport by designing the geometry of 1D nanodevie.

\section*{Acknowledgments}
This work is jointly supported by the National Key R\&D Program of China (Grant No. 2017YFA0303702 and No. 2016YFE0129300), the National Nature Science Foundation of China (Grants No. 51721001, No. 11625418, No. 11474158, No. 51472114, No. 11404157, No. 11475085, and No. 11535005). Y.-L. W. was funded by the Natural Science Foundation of Shandong Province of China (Grant No. ZR2017MA010), Linyi University (Grant No. LYDX2016BS135) and the China Postdoctoral Science Foundation (Grant No. 2017M611770).


\begin{thebibliography}{99}
\bibitem{Deshpande2008} V. V. Deshpande and M. Bockrath, Nat. Phys. \tb{4}, 314 (2008).
\bibitem{Mourik2012} V. Mourik, K. Zuo, S. M. Frolov, S. R. Plissard, E. P. A. M. Bakkers, and L. P. Kouwenhoven, Science \tb{336}, 1003 (2012).
\bibitem{HJensen1971} H. Jensen and H. Koppe, Ann. Phys. \tb{63}, 586-591(1971).
\bibitem{Costa1981} R. C. T. da Costa, Phys. Rev. A \tb{23}, 1982 (1981).
\bibitem{Jaffe2003} P. C. Schuster and R. L. Jaffe, Ann. Phys. \tb{307}, 132-143(2003).
\bibitem{Aoki2001} H. Aoki, M. Koshino, D. Takeda, H. Morise, and K. Kuroki, Phys. Rev. B \tb{65}, 035102 (2001).
\bibitem{Ortix2010} C. Ortix and J. van den Brink, Phys. Rev. B \tb{81}, 165419 (2010).
\bibitem{Campo2014} A. del Campo, M. G. Boshier and A. Saxena, Sci. Rep. \tb{4}, 5274 (2014).
\bibitem{Wang2016b} Y.-L. Wang, G.-H. Liang, H. Jiang, W.-T. Lu, and H.-S. Zong, J. Phys. D: Appl. Phys. \tb{49}, 295103(2016).
\bibitem{Wang2017} Y.-L. Wang, H. Jiang, and H.-S. Zong, Phys. Rev. A \tb{96}, 022116 (2017).
\bibitem{Liu2011} Q. H. Liu, C. L. Tong and M. M. Lai, J. Phys. A: Math. Theor. \tb{40}, 4161 (2007); Q. H. Liu, L. H. Tang, and D. M. Xun, Phys. Rev. A \tb{84}, 042101 (2011); Q. H. Liu, J. Math. Phys. \tb{54}, 122113 (2013); D. K. Lian, L. D. Hu, and Q. H. Liu, arXiv: 1701.08370.
\bibitem{Szameit2010} A. Szameit, F. Dreisow, M. Heinrich, R. Keil, S. Nolte, A. T\"{u}nnermann, and S. Longhi, Phys. Rev. Lett. \tb{104}, 150403 (2010).
\bibitem{Schmidt2015} R. Spittel, P. Uebel, H. Bartelt, and M. A. Schmidt, Opt. Express \tb{23}, 12174 (2015).
\bibitem{Ferrari2008} G. Ferrari and G. Cuoghi, Phys. Rev. Lett. \tb{100}, 230403(2008).
\bibitem{Jensen2009} B. Jensen and R. Dandoloff, Phys. Rev. A \tb{80}, 052109 (2009).
\bibitem{Ortix2011} C. Ortix and J. van den Brink, Phys. Rev. B \tb{83}, 113406 (2011).
\bibitem{Wang2014} Y.-L. Wang, L. Du, C.-T. Xu, X.-J. Liu, and H.-S. Zong, Phys. Rev. A \tb{90}, 042117 (2014).
\bibitem{Sun2009} N. Zhao, H. Dong, S. Yang, and C. P. Sun, Phys. Rev. B \tb{79}, 125440 (2009).
\bibitem{Brandt2015} F. T. Brandt and J. A. S\'{a}nchez-Monroy, EPL \tb{111}, 67004 (2015).
\bibitem{Taira2010b} H. Taira and H. Shima, J. Phys. A: Math. Theor. \tb{43}, 354013 (2010).
\bibitem{Chang2013} J.-Y. Chang, J.-S. Wu, and C.-R. Chang, Phys. Rev. B \tb{87}, 174413 (2013).
\bibitem{Ortix2015PRL} P. Gentile, M. Cuoco, and C. Ortix, Phys. Rev. Lett. \tb{115}, 256801 (2015).
\bibitem{Shikakhwa2017} M. S. Shikakhwa, Eur. Phys. J. Plus \tb{132}, 17 (2017).
\bibitem{Ortix2015} C. Ortix, Phys. Rev. B \tb{91}, 245412(2015).
\bibitem{Taira2010} H. Taira and H. Shima, J. Phys.: Condens. Matter \tb{22}, 075301 (2010).
\bibitem{Maraner1995} P. Maraner, J. Phys. A: Math. Gen., \tb{28}, 2939 (1995).
\bibitem{Encinosa2005} M. Encinosa, L. Mott, and B. Etemadi, Phys. Scr. \tb{72}, 13 (2005).
\bibitem{Jaffe1999} P. Ouyang, V. Mohta, and R. L. Jaffe, Ann. Phys. \tb{275}, 297-313 (1999).
\bibitem{Schmelcher2014} J. Stockhofe and P. Schmelcher, Phys. Rev. A \tb{89}, 033630 (2014).
\bibitem{Wang2016} Y.-L. Wang and H.-S. Zong, Ann. Phys. \tb{364}, 68-78 (2016).
\bibitem{Iwo2017} Iwo Bialynicki-Birula and Zofia Bialynicka-Birula, Phys. Rev. Lett. \tb{118}, 114801 (2017).
\bibitem{Stephen2017} Stephen M. Barnett, Phys. Rev. Lett. \tb{118}, 114802 (2017).
\bibitem{Chen2008} X.-S. Chen, X.-F. L\"{u}, W.-M. Sun, F. Wang, and T. Goldman, Phys. Rev. Lett. \tb{100}, 232002 (2008).
\bibitem{Wang2015} F. Wang, C.-W. Wong, X.-S. Chen, W.-M. Sun, and P.-M. Zhang, Few-Body Syst. \tb{56}, 249 (2015).
\bibitem{Neil2002} A. T. O'Neil, I. MacVicar, L. Allen, and M. J. Padgett, Phys. Rev. Lett. \tb{88}, 053601 (2002).
\bibitem{Zhang2005} S. Zhang and Z. Yang, Phys. Rev. Lett. \tb{94}, 066602 (2005).
\bibitem{Dick2006} R. Dick, Eur. Phys. J. B \tb{53}, 127 (2006).
\bibitem{Qi2009} X.-L. Qi and S.-C. Zhang, Phys. Rev. B \tb{79}, 235442 (2009).
\end{thebibliography}
\end{document}